\documentclass[english,aps,reprint]{revtex4}
\usepackage[T1]{fontenc}
\usepackage[latin9]{inputenc}
\usepackage{color}
\usepackage{float}
\usepackage{amsmath}
\usepackage{graphicx}
\usepackage{amssymb}
\usepackage{esint}

\newcommand{\be}{\begin{equation}}
\newcommand{\ee}{\end{equation}}
\newcommand{\bea}{\begin{eqnarray}}
\newcommand{\eea}{\end{eqnarray}}

\newcommand{\ba}{\begin{eqnarray}}
\newcommand{\ea}{\end{eqnarray}}
\newcommand{\ua}{\uparrow}
\newcommand{\da}{\downarrow}

\makeatletter

\@ifundefined{textcolor}{}
{%
 \definecolor{BLACK}{gray}{0}
 \definecolor{WHITE}{gray}{1}
 \definecolor{RED}{rgb}{1,0,0}
 \definecolor{GREEN}{rgb}{0,1,0}
 \definecolor{BLUE}{rgb}{0,0,1}
 \definecolor{CYAN}{cmyk}{1,0,0,0}
 \definecolor{MAGENTA}{cmyk}{0,1,0,0}
 \definecolor{YELLOW}{cmyk}{0,0,1,0}
 }

\makeatother

\makeatother

\usepackage{babel}

\makeatother

\usepackage{babel}

\makeatother

\usepackage{babel}

\begin{document}

\title{Fidelity spectrum and phase transitions of quantum systems 
 }

\author{P. D. Sacramento$^1$, N. Paunkovi\'c$^2$, V. R. Vieira$^1$}

\affiliation{$^1$ Departamento de F\'{\i}sica and CFIF, Instituto Superior T\'ecnico, TU Lisbon,
 Av. Rovisco Pais, 1049-001 Lisboa, Portugal}

\affiliation{$^2$ SQIG- Instituto de Telecomunica\c{c}\~oes, IST, TU Lisbon, Av. Rovisco Pais,
1049-001 Lisboa, Portugal
}

\begin{abstract}
%\textbf{preprint \today{}}\\
Quantum fidelity between two density matrices, $F(\rho_1,\rho_2)$ is usually defined as the trace
of the operator ${\cal F}=\sqrt{ \sqrt{\rho_1} \rho_2 \sqrt{\rho_1}}$. We study the logarithmic spectrum
of this operator, which we denote by {\it fidelity spectrum}, 
in the cases of the $XX$ spin chain in a magnetic field, a magnetic impurity
inserted in a conventional superconductor and a bulk superconductor at finite temperature. When the density
matrices are equal, $\rho_1=\rho_2$, the fidelity spectrum reduces to the entanglement spectrum.
We find that the fidelity spectrum can be a useful tool in giving a detailed 
characterization of different phases of many-body quantum systems.
\end{abstract}

\maketitle
\global\long\def\ket#1{\left| #1\right\rangle }

\global\long\def\bra#1{\left\langle #1 \right|}

\global\long\def\kket#1{\left\Vert #1\right\rangle }

\global\long\def\bbra#1{\left\langle #1\right\Vert }

\global\long\def\braket#1#2{\left\langle #1\right. \left| #2 \right\rangle }

\global\long\def\bbrakket#1#2{\left\langle #1\right. \left\Vert #2\right\rangle }

\global\long\def\av#1{\left\langle #1 \right\rangle }

\global\long\def\tr{\text{Tr}}

\global\long\def\im{\text{Im}}

\global\long\def\re{\text{Re}}

\global\long\def\sign{\text{sign}}

\global\long\def\Det{\text{Det}}

%\tableofcontents{}\newpage{}

\section{Introduction}
A quantum system in a pure state is described by a density matrix which
is just a projector onto that state. At zero temperature it is the projector
to the groundstate of the system. In general, the Hamiltonian of the system is
a function of some parameters which determine the groundstate. The quantum
fidelity between two states (for two sets of parameters) is, in this simple
case, the absolute value of the overlap between the groundstates for the two sets of parameters.
When the system is in a mixed state the density matrix is more complex. Typical
situations that lead to mixed states are: i) reduced density matrices where a
trace over some degrees of freedom is carried out, or ii) systems at
finite temperatures where the density matrix may be taken as the Boltzmann
factor over the energy eigenstates.

In general the quantum fidelity \cite{Wooters} between two states characterized by two density
matrices $\rho_1$ and $\rho_2$ may be defined as the trace of the fidelity operator,
${\cal F}$,
\be
F(\rho_1,\rho_2) = \mbox{Tr} {\cal F}= \mbox{Tr} \sqrt{ \sqrt{\rho_1} \rho_2 \sqrt{\rho_1}}.
\label{eq1}
\ee
One may also consider the spectrum of the fidelity operator ${\cal F}(\rho_1,\rho_2)$.
Its set of eigenvalues ${\lambda_i}$, which we denote {\it fidelity operator spectrum},
and ${-\ln \lambda_i}$, which we call the {\it fidelity spectrum}, may provide more information
as compared to the fidelity (its trace), in a way parallel to the extra information
provided by the entanglement spectrum \cite{Haldane}, as compared to the von Neumann entropy.

In the case of pure states 
$\rho_1 = |GS_1\rangle \langle GS_1|$ and $\rho_2 = |GS_2\rangle \langle GS_2|$
the fidelity is just the norm of the overlap 
\be
F(\rho_1,\rho_2) = |\langle GS_1 | GS_2 \rangle |.
\ee
If the two states are the same it is just the normalization of the state (taken as 1).
In the case of two equal mixed states the fidelity is just
\be
F(\rho,\rho)=\mbox{Tr} \rho = 1
\ee
and the operator ${\cal F}$ in this case has a set of eigenvalues, ${\lambda_i}=\lambda_i^{\rho}$, such that
$-\ln \lambda_i$ is called the entanglement spectrum,
and has received considerable attention lately \cite{Haldane}. 

In this
work we will analyse the fidelity spectrum for several physical systems paying
particular attention to the vicinity of quantum phase transitions (QPT) as well as
the properties characterizing their quantum phases.

The interplay between quantum information and condensed matter physics has been
extensively considered using entanglement as a measure for the behavior of many body
systems \cite{Amico}.
The distinguishability between states has been used as a possible criterion to study
quantum phase transitions \cite{Zanardi}.
By its own nature, fidelity between pure groundstates signals a change of state as one approaches a quantum
phase transition \cite{Gu}. The fidelity between mixed states has also been
used as a signature of quantum phase transitions \cite{Zhou,us} and to distinguish
between different states of matter at finite temperatures \cite{BCS}.
A standard measure of entanglement in a system is the von Neumann entropy.
However, as argued in Ref. \cite{Haldane}, more information about a mixed state
is obtained if the entanglement spectrum is analysed.
Considering reduced density matrices where part of the degrees of freedom are
integrated over, such as dividing the system in real space into two parts $A$ and
$B$, it was shown in the context of the quantum Hall effect \cite{Haldane, Bernevig} and
in the context of coupled spin chains \cite{Poilblanc}, that the groundstate entanglement
spectrum of $A$ contains information about excited energy states of the frontier
of the subsystem $A$.
In particular, in the quantum Hall effect the entanglement spectrum of the bulk system 
has a low-lying structure of levels that matches the edge states, and in the case of
two coupled Heisenberg spin chains, considering the subsystem $A$ as one of the chains, the
entanglement spectrum has a structure that matches the energy excitations of a single
Heisenberg chain. Other partitionings of the system have been proposed that lead
to further information \cite{Bernevig2} and considering a partitioning in momentum
space it was shown that information about energy excitations of a single Heisenberg
chain is contained in the groundstate wave function through the entanglement spectrum
\cite{Arovas}.

In this work we consider the fidelity spectrum of various systems. 
While the entanglement spectrum has some relation to the energy spectrum of the edge states
or even bulk states, the fidelity spectrum contains information about which
eigenvalues have a larger contribution to the distinguishability between quantum states.
We start by considering
two systems at zero temperature: a magnetic impurity inserted in a conventional
superconductor and the $XX$ chain in a magnetic field. In the first case,
described in Section II, the magnetic impurity is coupled
through a spin interaction to the spin density of the conduction electrons, tuned by
a coupling $J$. As previously discussed \cite{Sakurai,Balatsky,us}, as the
coupling $J$ grows, the
system goes through a first order phase transition. At this point the system becomes
magnetized and various quantities such as local density of states, spin content,
gap function and quantum information measures can be used to detect this transition.
For instance, various entanglement measures \cite{Entanglement} and the partial state fidelity
itself \cite{us} have been used before. In the case of the $XX$ chain describing spins $1/2$ confined
to a plane ($xy$) and with a transverse magnetic field $h$, aligned along the $z$
direction (Section III), there is a quantum phase transition from a $XX$ phase, where the spins are
aligned in the $xy$ plane, if the magnetic field is small, and an Ising-like phase where
the spins point along the field direction, if the field is strong enough. Considering the
coupling between the spins as the energy scale, the second order transition occurs at the
point $h_c=1$. 
This transition is also signaled in various ways such as the 
decrease of the fidelity near the critical point \cite{Zanardi}. 
In Section IV we consider thermal states in a conventional superconductor.

\section{Magnetic impurity in a superconductor}

Consider first a classical spin immersed in a two-dimensional $s$-wave
conventional superconductor. We use a lattice description of the
system. In the center of the system, $i=l_c=(0,0)$, we place a
classical spin along the $z$-direction $\vec{S}=S \vec{e}_z$, with no loss
of generality.
The Hamiltonian of the system is given by
\begin{equation}
\label{Hamiltonian-def} H = -\sum_{\langle ij\rangle\sigma}
t_{ij}c_{i\sigma}^\dag c_{j\sigma}-\varepsilon_F\sum_{i\sigma}c_{i\sigma}^\dag c_{i\sigma}
+\sum_i\left(\Delta_i c_{i\ua}^\dag c_{i\da}^\dag+\Delta_i^\ast c_{i\da} c_{i\ua}\right)
-\sum_{\sigma\sigma^\prime}
J c_{l_c\sigma}^\dag\sigma^z_{\sigma\sigma^\prime} c_{l_c\sigma^\prime},
\end{equation}
where the first term describes the hopping of electrons between
different sites on the lattice, $\varepsilon_F$ is the chemical
potential, the third term is
the superconducting $s$-pairing with the site-dependent order
parameter $\Delta_i$, and the last term, with $J>0$, is the exchange interaction
between an electron at site $i=l_c$ and the magnetic impurity. The
hopping matrix is given by $t_{ij}=t\delta_{j,i+\delta}$ where
$\delta$ is a vector to a nearest-neighbor site. Note that both 
indices $i,j\in\{1,2,\ldots N\}$ specify sites on a
two-dimensional system ($N$ is the number of sites). We take
energy units in terms of $t$ ($t=1$), and choose $\varepsilon_F=-1$.

\begin{figure}[h]
\begin{centering}
\includegraphics[width=0.6\columnwidth]{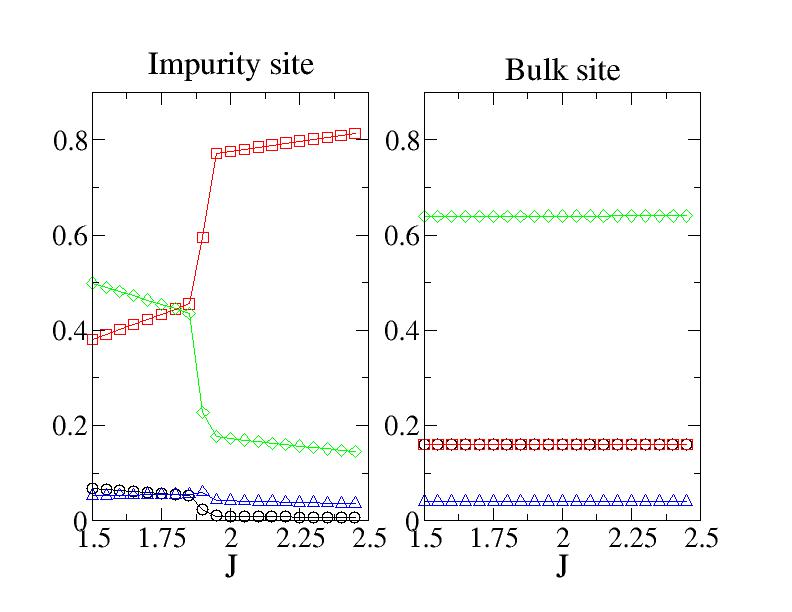}
\par\end{centering}

\caption{\label{fig1} (color online) Fidelity operator spectrum at the impurity site (left) and at a bulk site
(right) as a function of the spin coupling, $J$, where one density matrix is calculated
at $J$ and the other at $J+\delta J$, where $\delta J=0.05$. In black and red is the charge contribution and in
green and blue the spin contribution.
}
 
\end{figure}

\begin{figure}[h]
\begin{centering}
\includegraphics[width=0.45\columnwidth]{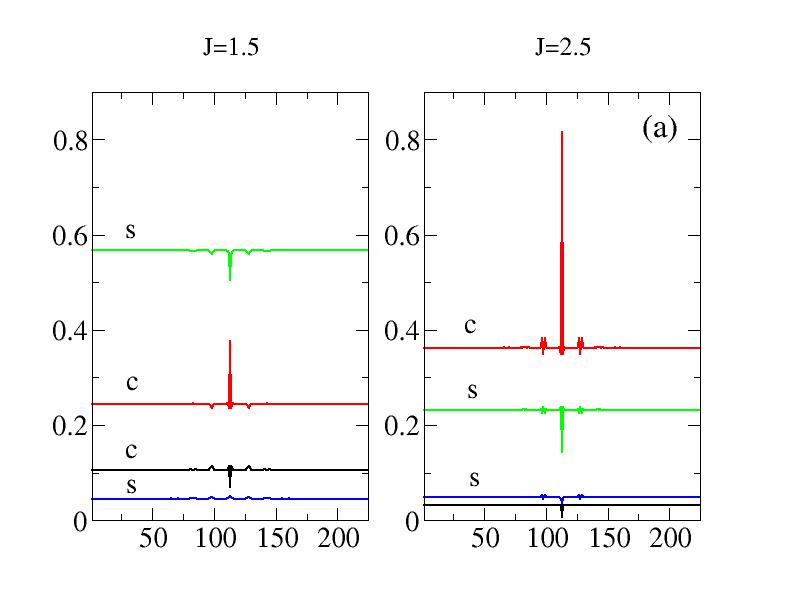}
\includegraphics[width=0.45\columnwidth]{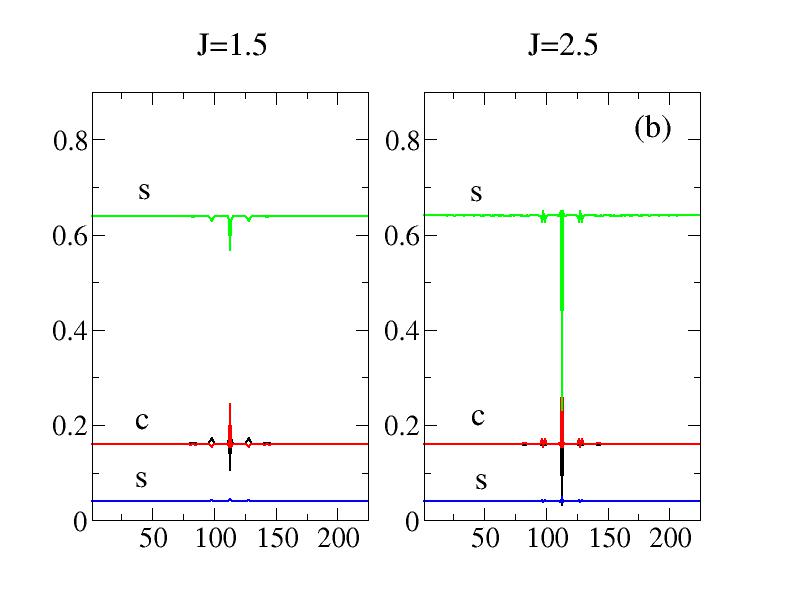}
\par\end{centering}

\caption{\label{fig2} (color online) Fidelity operator spectrum: charge
eigenvalues in black and red; spin eigenvalues in green and blue. System size is $15\times 15$. 
a) One of the sites is at the impurity, $i_1=l_c$, and the other site, $i_2$, is arbitrary. In the left panel 
$J_1=J_2=1.5$, and in the right panel $J_1=J_2=2.5$.
b) One of the sites, $i_1$, is a site in the bulk and the other site, $i_2$, is arbitrary.
The other parameters are the same as in a).
Recall that $J=1.5<J_c$ and $J=2.5>J_c$.
 }
 
\end{figure}

\begin{figure}[h]
\begin{centering}
\includegraphics[width=0.24\columnwidth]{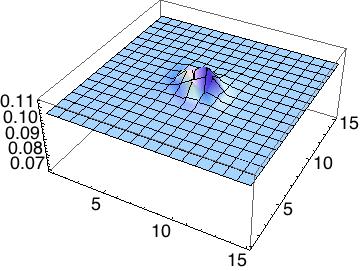}
\includegraphics[width=0.24\columnwidth]{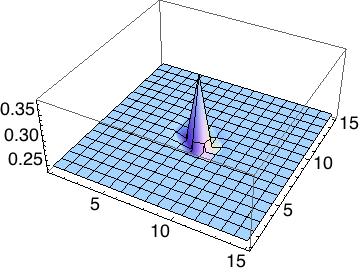}
\includegraphics[width=0.24\columnwidth]{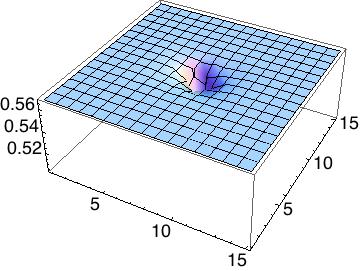}
\includegraphics[width=0.24\columnwidth]{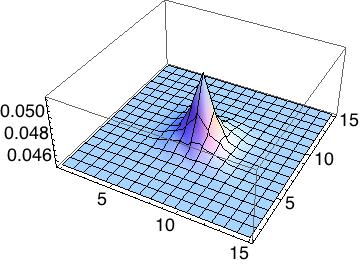}
\includegraphics[width=0.24\columnwidth]{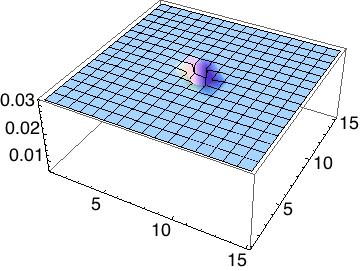}
\includegraphics[width=0.24\columnwidth]{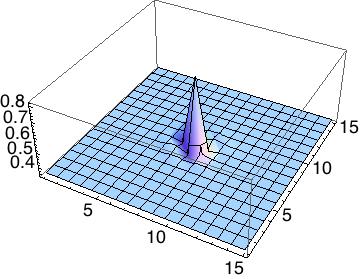}
\includegraphics[width=0.24\columnwidth]{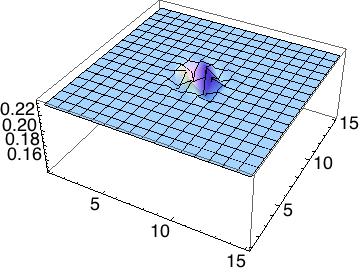}
\includegraphics[width=0.24\columnwidth]{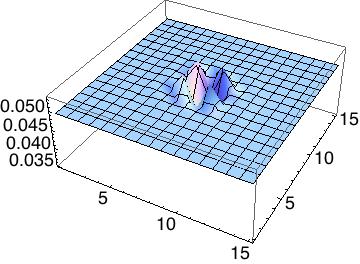}
\par\end{centering}

\caption{\label{fig3} (color online) Fidelity operator spectrum as a function of space. System size is $15\times 15$. 
Top panels $J=1.5$. Lower panels $J=2.5$.
From left to right, the first two panels are the charge eigenvalues
and the right two panels the spin eigenvalues, for lattice sites
$i_1=l_c$ and $i_2$ an arbitrary site.
}
 
\end{figure}

If we divide the whole
system in two subsystems, say $A$ and $B$, then the partial mixed
state, given by the reduced density operator $\rho_A$ for the
subsystem $A$, is defined as
\begin{equation}
\rho_A=\mbox{Tr}_B \rho,
\end{equation}
where $\mbox{Tr}_B[\cdot ]$ represents the partial trace evaluated
over the Hilbert space $\mathcal{H}_B$ of the subsystem $B$. 
We take $A$ to be one site, either the impurity site or an arbitrary site in
the bulk far from the impurity 
such that $\rho_1$ and $\rho_2$ of Eq. (\ref{eq1}) are one-site density matrices.

In many-body systems second quantization is the natural way to perform any calculation. 
The matrix elements of the
density matrix are simply defined in terms of correlation functions
of the whole system. For instance, in the case of the single-site
partial mixed states using local basis states
$\mathcal{B}=\{|0\rangle,|\!\ua\da\rangle,|\!\ua\rangle,|\!\da\rangle\}$,
which denote the four possible states --- unoccupied, double
occupied, single occupied with an electron with spin up and single
occupied with an electron with spin down, respectively --- it can be shown
that the corresponding density matrix reads as \cite{us}
\begin{equation}
\label{one-site_state} \rho_i=\left(\begin{array}{cccc}
\langle(1-n_\ua)(1-n_\da)\rangle &
\langle c_{\ua}^\dag c_{\da}^\dag\rangle & 0 & 0 \\
\langle c_{\da} c_{\ua}\rangle &
\langle n_\ua n_\da\rangle & 0 & 0 \\ 0 & 0 &
\langle n_\ua(1- n_\da)\rangle &
\langle c_{\da}^\dag c_{\ua}\rangle \\ 0 & 0 &
\langle c_{\ua}^\dag c_{\da}\rangle &
\langle(1-n_\ua) n_\da\rangle
\end{array}\right)_i,
\end{equation}
where the index $i$ denotes the site.
The spin and the charge parts decouple. The spin part couples
the two spin orientations (single occupied states) and the charge
part couples the empty and doubly occupied states. The diagonal
terms of the matrix describe the number of empty sites, the number
of doubly occupied sites, the number of spin up sites and the number
of spin down sites, respectively. The sum of the diagonal terms is
equal to $1$ due to normalization. The matrix is easily diagonalized
and the fidelity between two different one-site
states obtained straightforwardly.

We will consider two density matrices of the form $\rho_1(J_1;i_1)$,
and $\rho_2(J_2;i_2)$ for two values, $J_1$ and $J_2$, of the spin coupling $J$, and for
two sites, $i_1$ and $i_2$. 

In Fig. \ref{fig1} we show the fidelity operator spectrum as a function
of the coupling between the magnetic impurity and the electronic spin density
taking $J_1=J$, $J_2=J+0.05$ where $i_1=i_2=i$ is the impurity site or a bulk site.
The charge and spin parts separate and in the charge part there is
the empty and the doubly-occupied contributions and in the spin part
the spin up and spin down contributions. As the coupling $J$ grows there is a discontinuity
in the eigenvalues. This discontinuity is associated with the quantum phase transition
previously discussed. The sum of the four eigenvalues is the fidelity, as discussed
before. There is a discontinuity at the QPT both in the total fidelity,
in the charge and spin parts and in the individual eigenvalues as well.
As we can see, the discontinuities occur mainly in one of the charge
eigenvalues and in one of the spin eigenvalues. As one crosses the QPT
to a regime where the impurity captures one electron breaking a Cooper
pair we see that the main contribution to the discontinuity in the charge
part comes from the doubly-occupied states, where there is a significant 
increase at the QPT. In the same way there is a significant decrease
in the spin up eigenvalue, leading to a smaller spin contribution
beyond the QPT. For small values of $J$ there is a screening of
the perturbation induced by the magnetic impurity in the superconductor.
A small fidelity means a higher degree of distinguishability.
For small values of $J$ the spin down eigenvalue is small but the spin up
is still high. Beyond the QPT both contributions are small. This indicates
that the transition is mainly of spin character: as the 
fidelity tends asymptotically to 1 away from the QPT and the spin eigenvalues
are small, the charge
eigenvalues have to compensate, however, mainly through the doubly-occupied
contribution, as expected. Note that, in the bulk, the transition is quite
small. The physics is very local, centered around the impurity site.

In Fig. \ref{fig2} we consider that the coupling is fixed, but the two density
matrices are calculated at different sites $i_1$ and $i_2$: $\rho_1(J,i_1)$ and
$\rho_2(J,i_2)$.
The various cases are specified in the caption
of the figure. In these figures the horizontal axis is the
lattice site. For each site there are four eigenvalues. The central point
is the impurity site. The symmetrical peaks close to the central point
are the neighbors of the central point (please note that here we number the
lattice sites of the 2d system as a 1d system sequentially row by row; so the lattice
nearest neighbors in the x-direction are close by neighbors but the neighbors in the y direction
are far apart). The fluctuations around the central peak are of course
better seen in a 3d plot. This is shown in Fig. \ref{fig3} where we compare the
four eigenvalues with the results of Fig. \ref{fig1}a.
As discussed above the empty site and the spin down contributions are small
and do not change much as we change from $J=1.5$ (below the QPT) to
$J=2.5$ (above the QPT).

\section{$XX$ spin-$1/2$ chain in a transverse field}

The $XX$ spin $1/2$ model has been solved exactly via the Jordan-Wigner
transformation where it is reduced to a system of free spinless fermions
\cite{Mattis}. The correlation functions were also calculated \cite{Barouch}
as well as the reduced density matrix of a system of L contiguous spins
\cite{Latorre1, Latorre2, Jin}. The information theoretic approach in terms
fidelity, Fisher metric and Chernoff bound was applied to the $XY$ model in
\cite{Zanardi, Giorda, Cozzini, Zanardi_g1, Abasto, Garnerone, Dubail}.

The Hamiltonian we will consider here is of the form
\be
H = -\frac{1}{2} \sum_{l=0}^{N-1} \left( \frac{1+\gamma}{2} \sigma_l^x \sigma_{l+1}^x +
\frac{1-\gamma}{2} \sigma_l^y \sigma_{l+1}^y + h \sigma_l^z \right).
\ee
Here the spin operators are described by Pauli matrices, $h$ is the
transverse magnetic field and $\gamma$ is the anisotropy.
We will simplify and consider $\gamma=0$ and we take $0<h<1$.
We will consider a block of $L$ contiguous spins. The reduced density
matrix of the block can be written as \cite{Latorre1}
\be
\rho_A=\prod_{i=1}^L \left( \frac{1+\nu_i}{2} b_i^{\dagger} b_i +
\frac{1-\nu_i}{2} b_i b_i^{\dagger} \right).
\label{eq8}
\ee
The operators $b_i$ are spinless fermionic operators with a 2-state
space (the eigenvalues of the number operator $n_i=b_i^{\dagger} b_i$ are $1$
and $0$, corresponding to occupied or
empty state, respectively). The eigenvalues of the reduced density matrix
are $2^L$ in number as the result of the direct product of the
$i=1,\cdots,L$ subspaces. Defining two Majorana operators of the form
\bea
c_{2l-1} &=& \left( \prod_{n=1}^{l-1} \sigma_n^z \right) \sigma_l^x \nonumber \\
c_{2l} &=& \left( \prod_{n=1}^{l-1} \sigma_n^z \right) \sigma_l^y
\eea
in terms of the spin operators, it has been shown that
\be
\langle GS | c_m c_n | GS \rangle = \delta_{m,n} + i \left( B_L \right)_{mn}.
\ee
The matrix $B_L$ is written as 
\be
B_L = G_L \otimes 
\left(\begin{array}{cc}
0 & 1 \\
-1 & 0
\end{array}\right)
\end{equation}
with
\begin{equation}
G_L=\left(\begin{array}{cccccc}
g_0 & g_{-1} & . & . & . & g_{1-L} \\
g_1 & g_0 & . & . & . & . \\
. & . & . & . & . & . \\
. & . & . & . & . & . \\
. & . & . & . & . & . \\
g_{L-1} & . & . & . & . & g_0 \\
\end{array}\right).
\end{equation}
We have that $g_l=g_{-l}$, $g_0=(2 \varphi_c/\pi)-1$ and
$g_{l \neq 0}=(2/l\pi) \sin l \varphi_c$, where $\varphi_c=\arccos (h)$.
Defining new Majorana fermions through the transformation $d=V c$ and imposing that
\be
\langle GS | d_m d_n | GS \rangle = \delta_{m,n} + i \left( \tilde{B}_L \right)_{mn}
\ee
with
\be
\tilde{B}_L = V B_l V^T = \Omega \otimes \left(\begin{array}{cc}
0 & 1 \\
-1 & 0
\end{array}\right),
\end{equation}
where $\Omega$ is a diagonal matrix with $L$ diagonal elements $\nu_l$, leads
to the diagonal form Eq. (\ref{eq8}) of the reduced density matrix, $\rho_A$, having defined $L$ complex fermionic
fields like
\be
b_l = \frac{d_{2l}+i d_{2l+1}}{2}.
\ee
The transformation $V$ that block diagonalizes the problem depends on the Hamiltonian
parameters through the numbers $g_l$. 

In order to calculate the fidelity operator ${\cal F}=\sqrt{ \sqrt{\rho_1} \rho_2 \sqrt{\rho_1}}$
 one needs to consider the product of two
reduced density matrices for two different magnetic fields. Even though each of
the reduced density matrix can be diagonalized, the transformations needed to diagonalize
$\rho_1$ and $\rho_2$ are different. 
The diagonalization
of $\rho_1$ ($\rho_2$) is obtained introducing a matrix $V_1$ ($V_2$). To obtain the
spectrum of ${\cal F}$ we rewrite the expression for the diagonalized reduced density
matrix $\rho_2$ in terms of the fermionic operators of the matrix $\rho_1$ which leads to
$\rho_2 = e^{-H_2}$, where 
\be
H_2 = - 
\sum_{i=1}^L \sum_{j=1}^L \sum_{l=1}^L
\left( \ln \frac{1+\nu_{2,l}}{2} T_{l,i} T_{l,j} |1\rangle_i \langle 0|_i \otimes |0\rangle_j
\langle 1|_j +
 \ln \frac{1-\nu_{2,l}}{2} T_{l,i} T_{l,j} |0\rangle_i \langle 1|_i \otimes |1\rangle_j
\langle 0|_j \right).
\ee
Here
\be
T_{l,i} = \sum_{p=1}^L {\bar V}_{2,l,p} \left({\bar V}\right)^{-1}_{1,p,i}
\ee
where we defined ${\bar V}$ by
\be
V = {\bar V} \otimes \left(\begin{array}{cc}
0 & 1 \\
-1 & 0
\end{array}\right).
\end{equation}
The diagonal elements $\nu_{2,l}$ are the diagonal elements of $\Omega$ for the
magnetic field of $\rho_2$.
Diagonalizing $H_2$, we obtain $\rho_2$, expressed in the eigenbasis of $\rho_1$,
and obtain the spectrum of ${\cal F}$, as intended.
Note that this requires diagonalizing a $2^L \times 2^L$ matrix which is much
larger than the $L \times L$ matrix required for the entanglement entropy.

\begin{figure}[t]
\begin{centering}
\includegraphics[width=0.3\columnwidth]{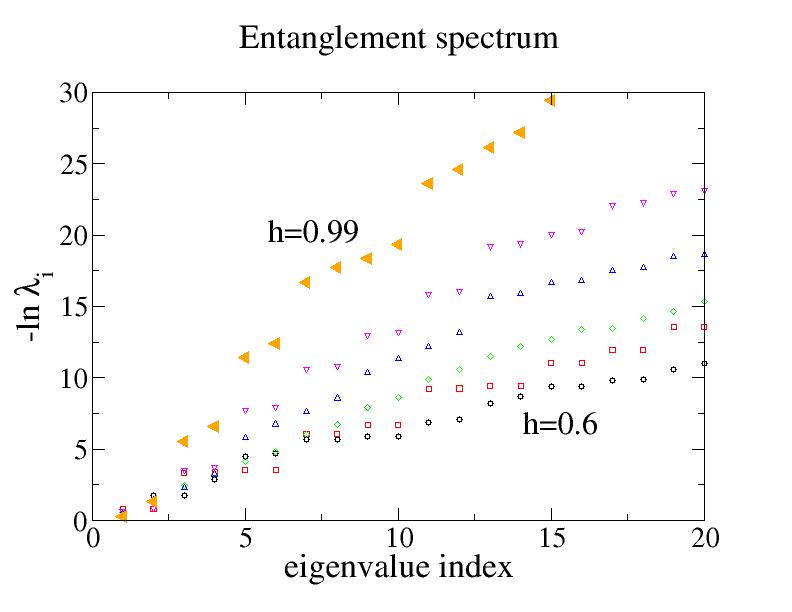}
\includegraphics[width=0.3\columnwidth]{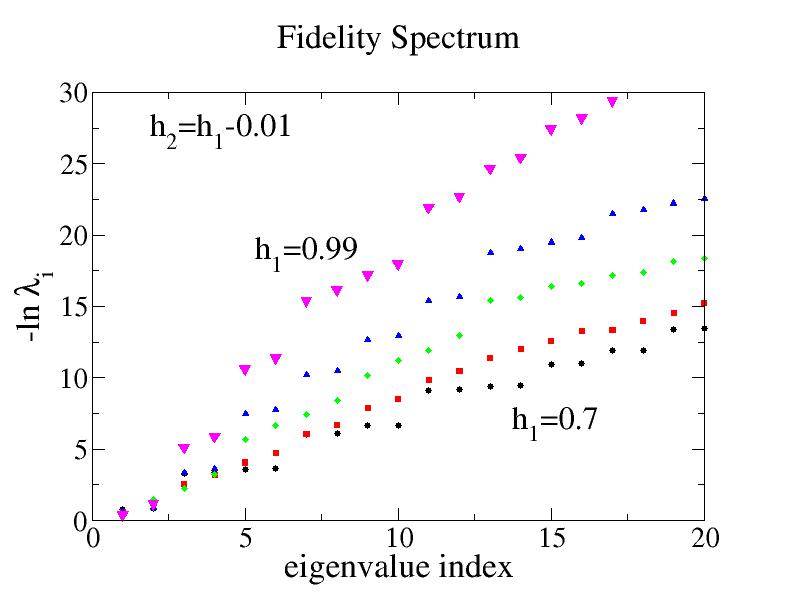}
\includegraphics[width=0.3\columnwidth]{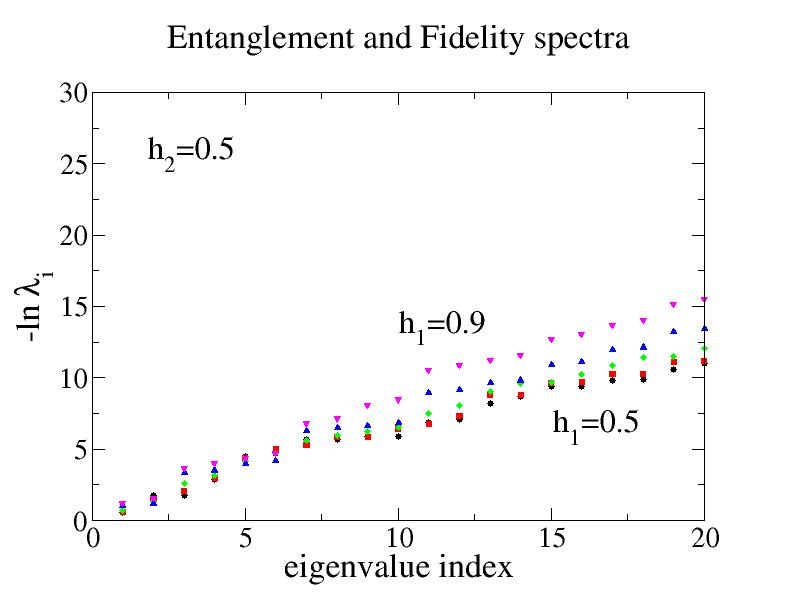}
\par\end{centering}

\caption{\label{fig4} (color online) Fidelity spectrum $-\ln \lambda_i$, for
system size $L=6$. 
Left: entanglement spectrum for different values of magnetic field.
From bottom to top $h=0.6,0.7,0.8,0.9,0.95,0.99$.
Middle: fidelity spectrum for $\delta h=0.01$ and for 
$h=0.7,0.8,0.9,0.95,0.99$.
Right: fidelity spectrum for pairs of the magnetic field as
$h_1=0.5$, $h_2=0.5,0.6,0.7,0.8,0.9$.
%, from down to up
}
 
\end{figure}

\begin{figure}[h]
\begin{centering}
\includegraphics[width=0.45\columnwidth]{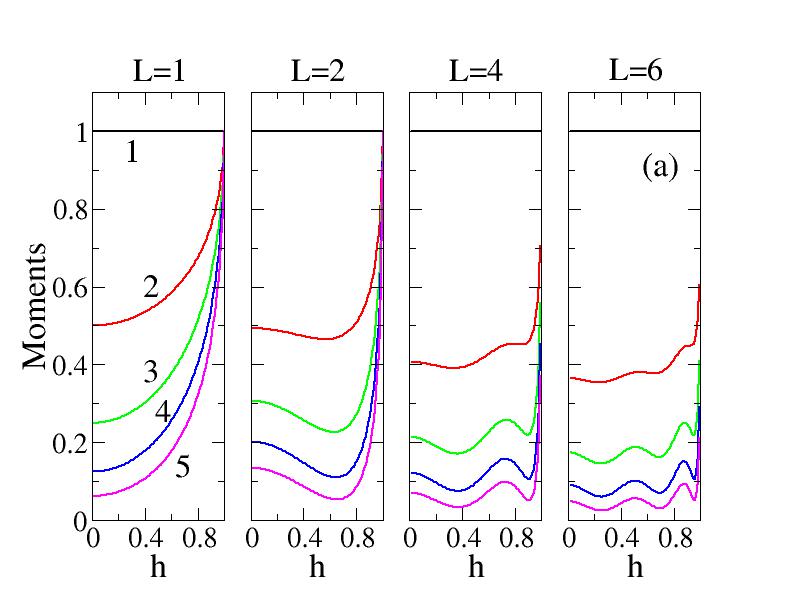}
\includegraphics[width=0.45\columnwidth]{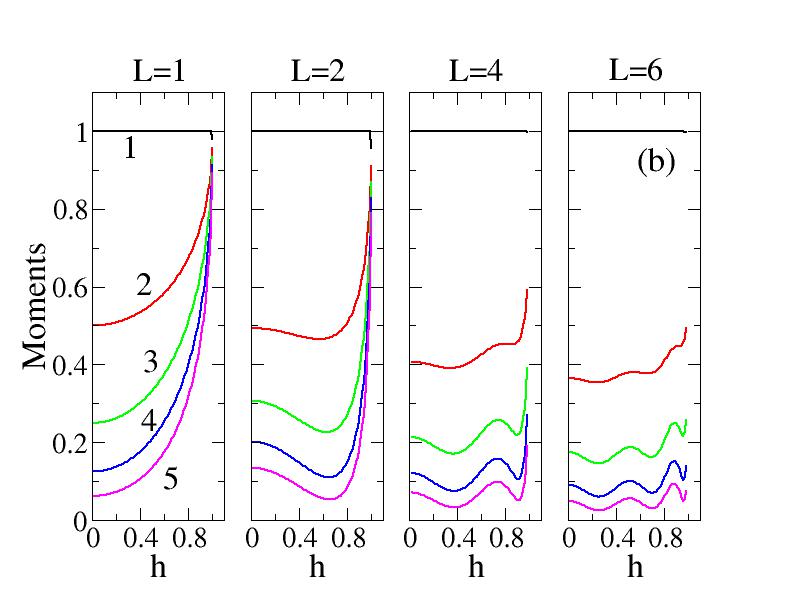}
\includegraphics[width=0.45\columnwidth]{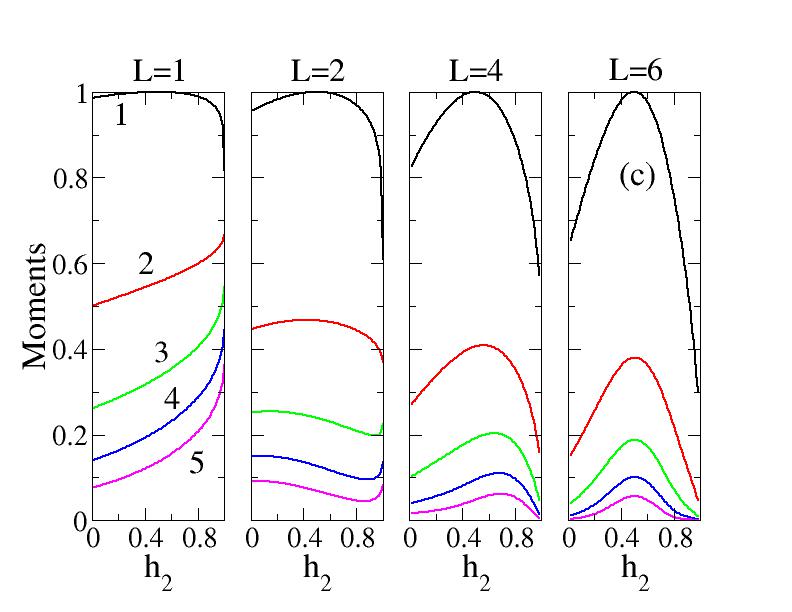}
\includegraphics[width=0.45\columnwidth]{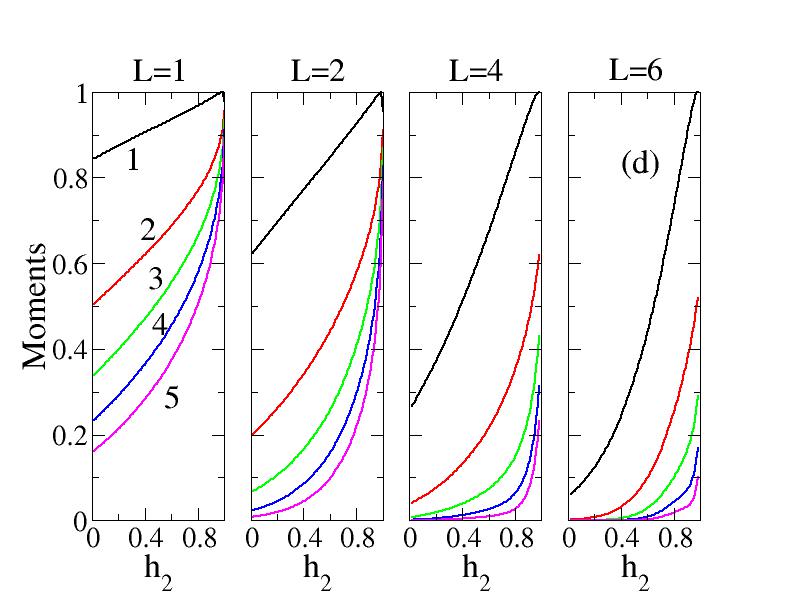}
\par\end{centering}

\caption{\label{fig5} (color online) First five moments of the entanglement operator (density matrix) and fidelity
operator spectra
for blocks of sizes $L=1,2,4,6$, as a function of magnetic field. In the top left panel we consider
the moments of the entanglement operator spectrum. In the top right panel we
consider the moments of the fidelity operator spectrum for two close-by values
of $h$. In the bottom left (right) panel we consider the fidelity operator spectra
for $h_1=0.5$ ($h_1=0.98$) and $h_2$ arbitrary.
%, respectively. 
}
 
\end{figure}

\begin{figure}[h]
\begin{centering}
\includegraphics[width=0.6\columnwidth]{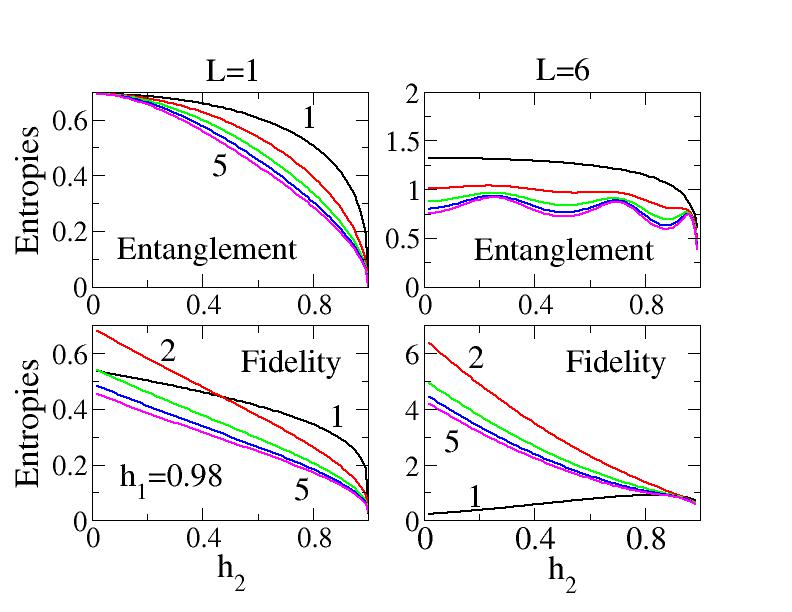}
\par\end{centering}

\caption{\label{fig6} (color online) R\'enyi and von Neumann entropies for $L=1$
(left) and $L=6$ (right), for
the first five moments, as a function of magnetic field.
In the top panels we consider the entanglement operator spectrum and in the bottom panels the
fidelity operator spectrum as a function of $h_2$, for $h_1=0.98$.
Note the decrease of the entropies close to the quantum critical
point, both for the entanglement case and the fidelity case. 
}
 
\end{figure}

\begin{figure}[h]
\begin{centering}
\includegraphics[width=0.45\columnwidth]{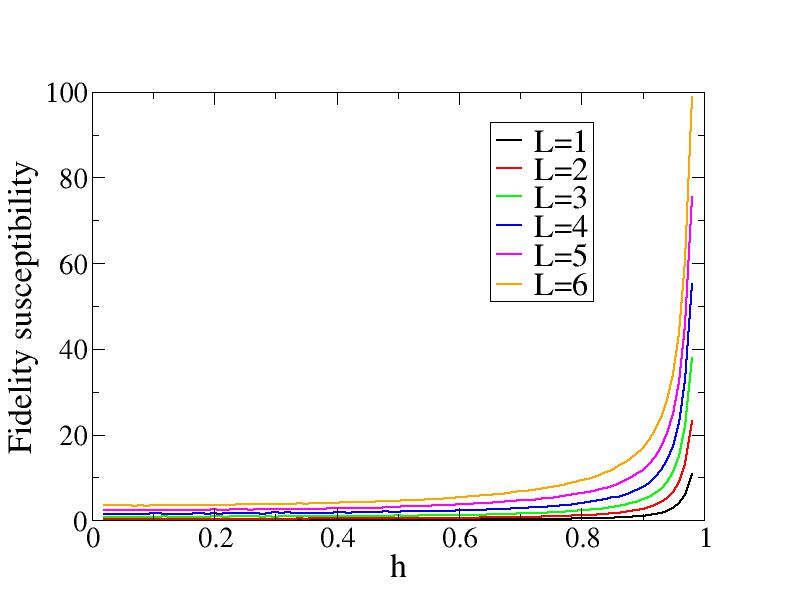}
\par\end{centering}

\caption{\label{fig7} (color online) Block fidelity susceptibility as a function of magnetic field
for 
%$L=1,2,3,4,5,6$ 
$L=1, \ldots, 6$ 
which shows the sharp increase
near the critical point. 
}
 
\end{figure}
In Fig. \ref{fig4} we show the fidelity spectrum 
$-\ln \lambda_i$. In the left panel we consider $\rho_1=\rho_2$ and the operator
${\cal F}$ is just the density matrix. Therefore, its logarithmic spectrum is the entanglement spectrum.
In the middle panel we consider the fidelity spectrum and in the right panel we compare the
entanglement spectrum with the fidelity spectrum.
As discussed before \cite{Arovas} there is no clear structure in the entanglement
spectrum since we are considering a real space block of spins.
The fidelity spectrum when we consider two very close values of the magnetic field
is very similar to the entanglement spectrum. In both cases as we increase
the magnetic field and approach the critical point that separates the
$XX$ phase from the Ising phase, the entanglement or fidelity spectrum
increases considerably. This implies that the fidelity operator spectrum
is decreasing fast. The same can be seen when we compare the fidelity
spectrum with the entanglement spectrum in the right panel. Here we are
considering two values of the magnetic field $h_1$ and $h_2$ associated
respectively with $\rho_1$ and $\rho_2$ that differ by a finite amount.
Therefore there is a significant difference as the difference between
$h_1$ and $h_2$ increases, even though we are far from the critical
regime. The further the two points are the smaller the fidelity
and the fidelity operator eigenvalues, $\lambda_i$, should be.

In order to analyse the spectrum we calculate the moments of the distribution
of the eigenvalues \cite{Pollmann} defined as
\be
M_n = \sum_{i=1}^{2^L} \lambda_i^n.
\ee
The moment of order $n=1$ is the fidelity. We also calculate the von Neumann
entropy of the fidelity
\be
S_1= -\sum_{i=1}^{2^L} \lambda_i \ln \lambda_i 
\ee
and the R\'enyi entropies
\be
S_n = \frac{1}{1-n} \ln M_n.
\ee
Note that in the case of the entanglement ($\rho_1=\rho_2$) the von Neumann
entropy can also be obtained as
\be
S_1 = -\sum_{i=1}^{L} \left( \frac{1+\nu_i}{2} \ln \frac{1+\nu_i}{2} +
\frac{1-\nu_i}{2} \ln \frac{1-\nu_i}{2} \right)
\ee
which only involves a sum with $L$ terms.

\begin{figure}[t]
\begin{centering}
\includegraphics[width=0.3\columnwidth]{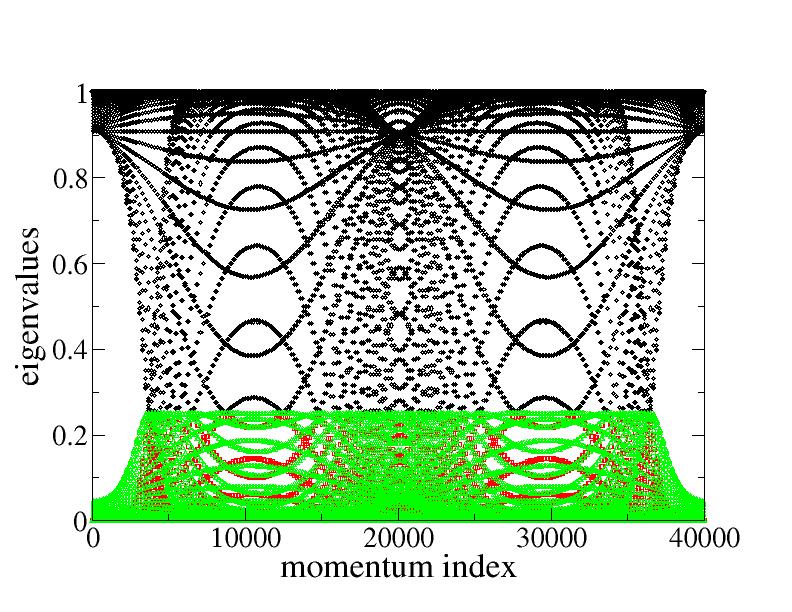}
\includegraphics[width=0.3\columnwidth]{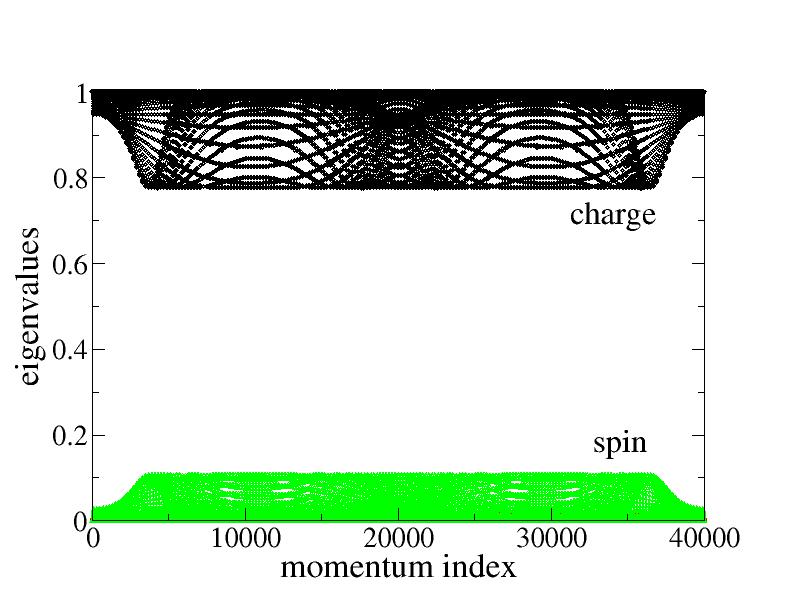}
\par\end{centering}

\caption{\label{fig8} (color online) Exponential entanglement spectrum for a bulk superconductor
as a function of momentum, labeled sequentially row by row in the Brillouin zone, in the normal phase
($\Delta=0$, left panel), and in the superconducting phase ($\Delta \neq 0$,
right panel), at temperature $T$. Note that in the right panel the lowest charge eigenvalue
is smaller than the spin eigenvalues. 
}
 
\end{figure}

\begin{figure}[t]
\begin{centering}
\includegraphics[width=0.3\columnwidth]{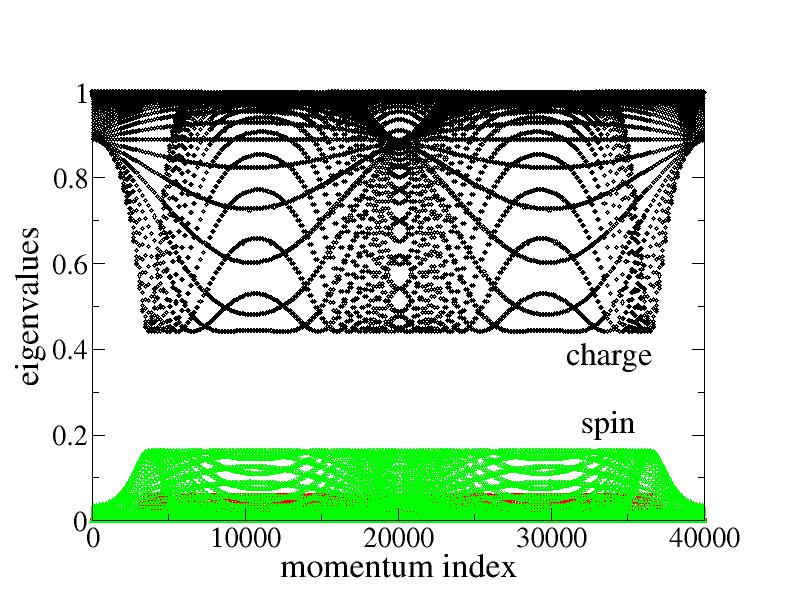}
\includegraphics[width=0.3\columnwidth]{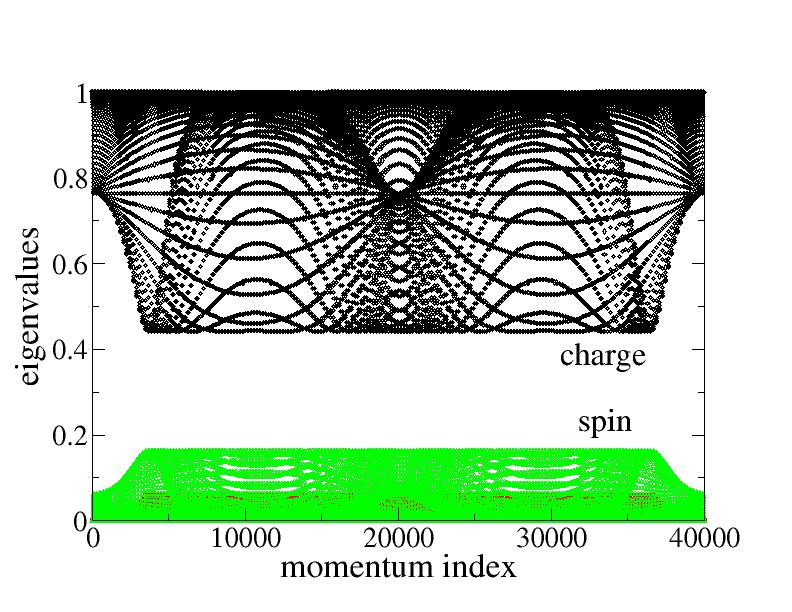}
\par\end{centering}

\caption{\label{fig9} (color online) Fidelity operator spectrum for a bulk superconductor
as a function of momenta, between the normal phase and the superconducting
phase, at the same temperature (left) and at different temperatures (right). 
}
 
\end{figure}

In Fig. \ref{fig5} we present the first five moments for several choices of pairs
of magnetic fields $(h_1,h_2)$ and for several block sizes, $L$.
In the case of the entanglement spectrum, when $h_1=h_2$, the first moment is just
the trace of the density matrix (top left panel). When $h_1 \neq h_2$ it is the fidelity and
therefore it is very close to 1 except in the vicinity of the phase transition
where we consider $h_2=h_1-\delta h, \delta h=0.01$ (top right panel). 
The higher momenta are very similar between the
entanglement spectrum and the fidelity spectrum, in this case. 
A detailed study of the moments and R\'enyi entropies has been carried
out in Ref. \cite{who}.
Considering two values of the magnetic fields further away from each other the moments change
in structure. Fixing for instance $h_1=0.5$ (bottom left panel) or 
$h_1=0.98$ (bottom right panel) and varying $h_2$ we
find that when $h_2$ crosses $h_1$ there is a sharp increase which equals
one for the first moment (trivially since for this case the first moment is
the trace of the density matrix). This maximum is also observed in the higher
moments of the spectrum. A similar information can be obtained from the R\'enyi
entropies. As explained above, $S_1$ is just the von Neumann entropy and the
other entropies are proportional to the logarithms of the moments of the
spectrum. In Fig. \ref{fig6} we compare various cases for $L=1,6$. As the magnetic
field approaches the critical point we find that the various entropies have
a minimum near the critical point showing that the partial state density
matrix signals the QPT, as previously obtained. In the lower panels we consider
the R\'enyi entropies associated with the fidelity operator. The structure is in
general more complicated. Fixing $h_1=0.98$ close to the critical point, as
$h_2$ approaches $h_1$ the entropies tend to those corresponding to the
entanglement spectrum. Far from this point the entropies differ considerably
and depend strongly on the block size. For instance for $L=1$ they are of the
order of the entropies for the entanglement spectrum but for $L=6$ the entropies
are considerably higher, except the von Neumann entropy.
Also, note that the R\'enyi entropy $S_1$
in the fidelity case has a depression at small magnetic fields.

We also calculate the fidelity susceptibility, introduced in 
\cite{You} and \cite{Zanardi_g1}, 
%\cite{You, Zanardi_g1}, 
with its geometrical meaning discussed in 
\cite{Zanardi_g1,Zanardi_g2}. Global fidelity susceptibility for the $XY$ model was discussed in
\cite{Zanardi,Zanardi_g1}. In the present paper was calculate the fidelity
susceptibility associated to reduced density matrices of blocks of spins. It is
defined as
\bea
\chi_F &=& \sum_{i=1}^{2^L} \chi_{F,i} \nonumber \\
\chi_{F,i} &=& \frac{\partial^2 \lambda_i}{\partial (\delta h)^2}.
\eea

In Fig. \ref{fig7} we present the fidelity susceptibility associated with
blocks of different sizes as a function of the magnetic field. The divergence
of the susceptibility is clearly seen as we approach the QPT.

\section{Thermal states of a conventional BCS superconductor}

In this section we consider a conventional $s$-wave superconductor at finite temperature
described by the effective mean-field BCS Hamiltonian
\begin{equation}
\label{bcs-hamiltonian}
H_{BCS}^{eff} = \sum_{k} \varepsilon_k(n_{k\ua}+n_{-k\da}) - 
\sum_k(\Delta_k c_{k\ua}^\dag c_{-k\da}^\dag + \Delta^\ast_k 
c_{-k\da} c_{k\ua} - \Delta^\ast_k \langle c_{-k\da} c_{k\ua} \rangle ),
\end{equation}
with $\Delta_k=- V \langle c_{-k\da} c_{k\ua} \rangle$, 
where the lattice-mediated pairing interaction is constant and
non-vanishing between electrons around the Fermi level only.
The density matrix is given by \cite{BCS}
\begin{equation}
\rho = \frac{1}{Z}e^{-(H_{BCS}^{eff}-\mu N)/T} = \frac{e^{\sum_k\vec{\tilde{h}}_k \vec{T}_k + K}}{\mbox{Tr}[e^{\sum_k\vec{\tilde{h}}_k 
\vec{T}_k + K}]} = \frac{\prod_k e^{\vec{´\tilde{h}}_k
\vec{T}_k}}{\prod_k\mbox{Tr}[e^{\vec{\tilde{h}}_k \vec{T}_k}]},
\end{equation}
where $T$ is the temperature, $\vec{\tilde{h}}_k =
(\tilde{h}_k^{+},\tilde{h}_k^{-},\tilde{h}_k^0) =
(2\Delta^\ast_k/T,2\Delta_k/T,-2\bar{\varepsilon}_k/T)$,
$\vec{T}_k = (T_k^{+},T_k^{-},T_k^0)$,
$K=-1/T\sum_k (\bar{\varepsilon}_k + \Delta^\ast_k b_k)$ and
$\bar{\varepsilon}_k=\varepsilon_k-\mu$. The norms of the vectors
$\vec{\tilde{h}}_k$ are given by $\tilde{h}_k = 2 E_k/T$, with $E_k =
\sqrt{\bar{\varepsilon}_k^2+\left|\Delta_k\right|^2}$. 
The coefficients $\vec{\tilde{h}}_k =
\vec{\tilde{h}}_k(T,V)$ are functions of both the coupling
constant $V$ and the temperature $T$,
through the gap parameters $\Delta_k = \Delta_k(T,V)$ and the
chemical potential $\mu$. 
By $n_{k\sigma} = c^\dag_{k\sigma}c_{k\sigma}$ we
denote the one-particle number operators, while by
$b^\dag_k=c^\dag_{k\ua}c^\dag_{-k\da}$ and
$b_k=c_{-k\da}c_{k\ua}$.
The $T_k$ operators are given by $T_k^{+} = b^\dag_k$,
$T_k^{-} = b_k$ and $2T_k^0 + 1 = (n_{k\ua}+n_{-k\da})$
and form a $\mbox{su}(2)$ algebra.

The fidelity is given by
\begin{eqnarray}
F(\rho_a, \rho_b) = \mbox{Tr}
[(\rho_a^{1/2}\rho_b\rho_a^{1/2})^{1/2}] = \frac{\mbox{Tr}[(\prod_k
e^{\frac{\vec{a}_k}{2} \vec{T}_k} e^{\vec{b}_k \vec{T}_k}
e^{\frac{\vec{a}_k}{2} \vec{T}_k})^{1/2}]}{\prod_k (\mbox{Tr}[e^{\vec{a}_k
\vec{T}_k}] \mbox{Tr}[e^{\vec{b}_k \vec{T}_k}])^{1/2}}
= \mbox{Tr} \left[ \prod_k \left( {\cal F}_k \right)^{1/2} \right].
\end{eqnarray}
As for every $k$ the operators $\vec{T}_k$ form a
$\mbox{su}(2)$ algebra, and therefore by exponentiation define a Lie
group, we can write $e^{\frac{\vec{a}_k}{2} \vec{T}_k}
e^{\vec{b}_k \vec{T}_k} e^{\frac{\vec{a}_k}{2}
\vec{T}_k} = e^{2\vec{c}_k \vec{T}_k}$.
For each value of the momentum we get a 4-dimensional space respecting to
momentum states that are empty, doubly occupied or singly occupied by a spin
up or a spin down electron. The space is therefore of the type
$\mathcal{B}=\{|0\rangle,|\!\ua\da\rangle,|\!\ua\rangle,|\!\da\rangle\}$,
similarly to the problem of the impurity in a superconductor.
The fidelity operator is then easily diagonalized in this $4\times 4$ subspace.
We will study the possible eigenvalues of the fidelity operator
for each momentum. As before, the charge and spin parts separate. Moreover, in
this problem the two spin components are degenerate. Therefore it is enough
to look at three eigenvalues (two for the charge part and one for the spin part).
The results are presented in the Figs. 8-12 and discussed in more detail below.

\begin{figure}[t]
\begin{centering}
\includegraphics[width=0.3\columnwidth]{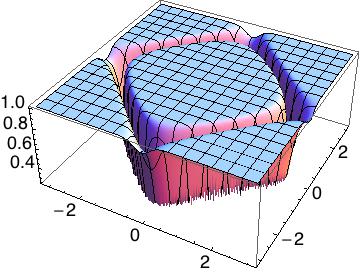}
\includegraphics[width=0.3\columnwidth]{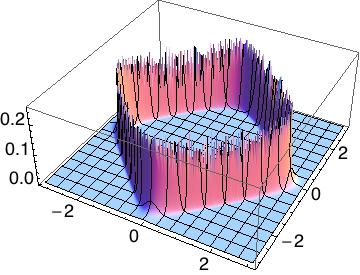}
\includegraphics[width=0.3\columnwidth]{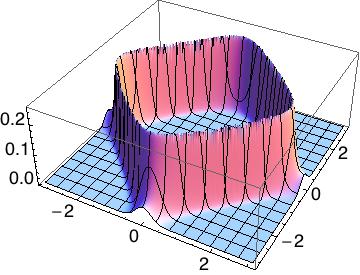}
\includegraphics[width=0.3\columnwidth]{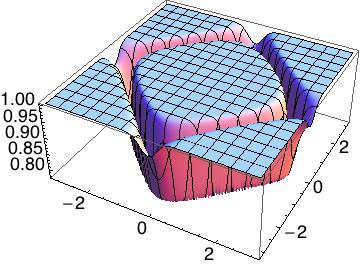}
\includegraphics[width=0.3\columnwidth]{fig10d.jpg}
\includegraphics[width=0.3\columnwidth]{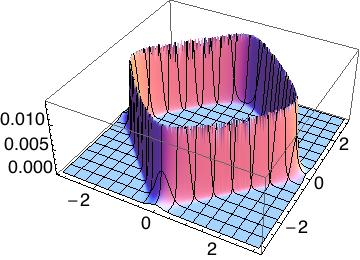}
\par\end{centering}

\caption{\label{fig10} (color online) Top (bottom) panels: normal (superconducting) phase 
exponential entanglement spectrum. 
In the three panels we show three eigenvalues since the two spin eigenvalues are degenerate.
}
 
\end{figure}

\begin{figure}[t]
\begin{centering}
\includegraphics[width=0.24\columnwidth]{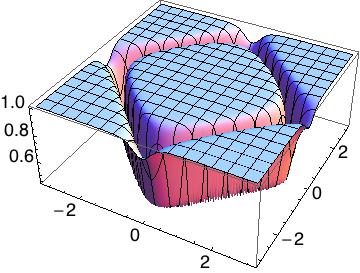}
\includegraphics[width=0.24\columnwidth]{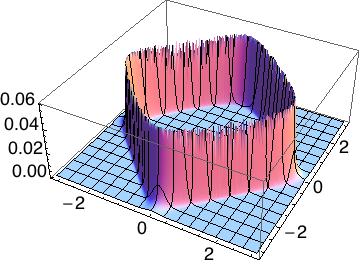}
\includegraphics[width=0.24\columnwidth]{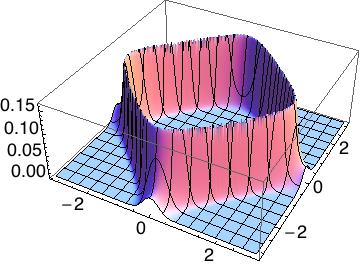}
\includegraphics[width=0.24\columnwidth]{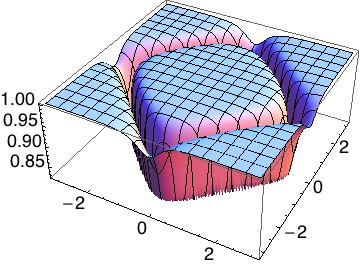}
\par\end{centering}

\caption{\label{fig11} (color online) Fidelity operator spectrum between superconducting and normal phases. In the first 3 panels
we show the 3 eigenvalues and in the last one the fidelity (smaller than 1 close to the
Fermi surface). Note that the temperature is the same, with $\Delta$ either finite (superconducting phase) or zero
(normal phase).
}
 
\end{figure}

The fidelity operator for each momentum value (denoted {\it k-fidelity operator}) is of the form ${\cal F}_k=\sqrt A_k$, where
\begin{equation}
\label{onesite_state} A_k = \frac{1}{D_k} \left(\begin{array}{cccc}
\alpha_k & \beta_k & 0 & 0 \\
\beta_k & \gamma_k & 0 & 0 \\
0 & 0 & 1 & 0 \\
0 & 0 & 0 & 1 \\
\end{array}\right).
\end{equation}
The matrix elements are given by
\bea
\alpha_k &=& \cosh E_k^a/T \cosh E_k^b/T +\sinh E_k^a/T \sinh E_k^b/T 
\frac{\Delta_k^a \Delta_k^b + \bar{\epsilon}_k^a \bar{\epsilon}_k^b}{E_k^a E_k^b} \nonumber \\
&+& \sinh E_k^a/T \cosh E_k^b/T \frac{\bar{\epsilon}_k^a}{E_k^a} 
+ \sinh E_k^b/T \frac{\bar{\epsilon}_k^b}{E_k^b} \nonumber \\
&+& \left( \cosh E_k^a/T-1 \right) \sinh E_k^b/T 
\frac{\Delta_k^a \Delta_k^b + \bar{\epsilon}_k^a \bar{\epsilon}_k^b}{E_k^a E_k^b}
\frac{\bar{\epsilon}_k^a}{E_k^a} \\
\gamma_k &=& \cosh E_k^a/T \cosh E_k^b/T +\sinh E_k^a/T \sinh E_k^b/T 
\frac{\Delta_k^a \Delta_k^b + \bar{\epsilon}_k^a \bar{\epsilon}_k^b}{E_k^a E_k^b} \nonumber \\
&-& \sinh E_k^a/T \cosh E_k^b/T \frac{\bar{\epsilon}_k^a}{E_k^a} 
- \sinh E_k^b/T \frac{\bar{\epsilon}_k^b}{E_k^b} \nonumber \\
&-& \left( \cosh E_k^a/T-1 \right) \sinh E_k^b/T 
\frac{\Delta_k^a \Delta_k^b + \bar{\epsilon}_k^a \bar{\epsilon}_k^b}{E_k^a E_k^b}
\frac{\bar{\epsilon}_k^a}{E_k^a} \\
\beta_k &=& \sinh E_k^a/T \cosh E_k^b/T \frac{\Delta_k^a}{E_k^a}
+ \sinh E_k^b/T \frac{\Delta_k^b}{E_k^b} \nonumber \\
&+& \left( \cosh E_k^a/T-1 \right) \sinh E_k^b/T 
\frac{\Delta_k^a \Delta_k^b +\bar{\epsilon}_k^a \bar{\epsilon}_k^b}{E_k^a E_k^b}
\frac{\Delta_k^a}{E_k^a}
\eea
and
\be
D_k = 2 \left( 1+\cosh E_k^a/T \right) 2 \left( 1+\cosh E_k^b/T \right).
\ee
The eigenvalues of the $k$-fidelity operator, ${\cal F}_k$ are therefore of the form
\begin{equation}
\label{onesite_state} \frac{1}{\sqrt{D_k}} \left(\begin{array}{cccc}
\eta_+^k & 0 & 0 & 0 \\
0 & \eta_-^k & 0 & 0 \\
0 & 0 & 1 & 0 \\
0 & 0 & 0 & 1 \\
\end{array}\right),
\end{equation}
where
\be
\eta_{\pm}^k = \frac{1}{2} \left[ (\alpha_k+\gamma_k) \pm
\sqrt{ (\alpha_k-\gamma_k)^2 + 4 \beta_k^2} \right].
\ee

We will be interested in situations where $\rho_1$ and $\rho_2$ correspond
to points in parameter space, which we choose to be the temperature, $T$, and the
gap function, $\Delta_k$, that are far apart and may be in the same or different
thermodynamic phases.

In Fig. \ref{fig8} we present the $k$-fidelity operator spectrum (of the operator ${\cal F}_k$)
for the case when the two density matrices are equal (``exponential entanglement
spectrum'')
where we compare the system in the normal phase (left) with the superconducting phase (right).
The horizontal axis is an index over the eigenvalues and in the vertical axis
we plot ${\lambda_k}$. We only plot three eigenvalues because the spin
eigenvalues are degenerate. For each label the sum over the four eigenvalues is 1 due to normalization.
In both phases the higher eigenvalue is the charge eigenvalue corresponding to
empty sites (this will be discussed later on). In the normal phase the lowest
eigenvalues merge into the higher eigenvalues but in the superconducting phase the
energy gap is clearly visible. Note that the eigenvalues are now labeled 
by the momentum. There is no partitioning of the system in real space but
there is a partitioning of the system in momentum space (since the system
can be block diagonalized). Recall however that here the mixed state originates
in the thermal states. We stress that we are not plotting the fidelity
operator, ${\cal F}=\prod_k {\cal F}_k$, eigenvalues.
In Fig. \ref{fig9} we consider two different density matrices where we plot the
$k$-fidelity operator
%, ${\cal F}_k$, 
spectrum where one of the density matrices
corresponds to a point in phase space in the normal phase and the other in
the superconducting phase. In the left panel the temperatures, $T_a,T_b$, are the same
(this can be obtained for instance considering two coupling constants)
and in the right panel the temperatures, $T_a,T_b$, are different. In the first case
the gap is still clearly visible. When the temperatures are different the gap
remains the same. There is a small decrease of the highest charge eigenvalue
that can be traced to the vicinity of the Fermi surface, as shown ahead.

In order to understand the spectrum in greater detail we consider the various eigenvalues in momentum
space.
In Fig. \ref{fig10} we consider the system in the normal phase and in the superconducting phase
for two equal density matrices $\rho_a=\rho_b$ and plot the
three eigenvalues of the $k$-fidelity operator as a function of momentum, for a 2d system. 
The fidelity is one for all momenta since $\mbox{Tr} \rho_a=\mbox{Tr}\rho_b=1$.
The depression in the highest eigenvalue (left panels) marks clearly the Fermi surface.
For momenta larger than the Fermi momentum there are no electrons (except
for thermal excitations contained in the Fermi function). So the eigenvalue
corresponding to empty states is 1. The other eigenvalues are close to zero
outside the Fermi surface. Due to the particle-hole transformation of the
Bogoliubov transformation, the empty site eigenvalue (corresponding to
doubly occupied sites in terms of the electrons) is also close to 1 inside
the Fermi surface. Accordingly, the other eigenvalues are also close to
zero inside the Fermi surface. The noticeable features are therefore
close to the Fermi surface. A similar structure is observed in the
superconducting phase. There is a slight change close to the Fermi surface
which is due to the opening of the superconducting gap. The amplitude of
the doubly occupied and spin eigenvalues are smaller in this case since
the decrease of the first eigenvalue is smaller along the Fermi surface.

In Fig. \ref{fig11} we show the fidelity operator spectrum for the case
when one density matrix is in the normal phase and the other corresponds to a quantum
state in the superconducting phase. The last panel shows the total fidelity.
It is significantly decreased around the Fermi surface where the difference
between the normal phase and the superconducting phase is larger due to the
pairing and opening of the gap. Note that the highest eigenvalue has a structure
that strongly resembles the total fidelity. 

\begin{figure}[t]
\begin{centering}
\includegraphics[width=0.3\columnwidth]{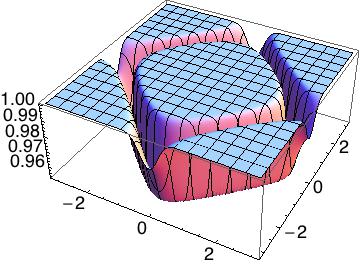}
\includegraphics[width=0.3\columnwidth]{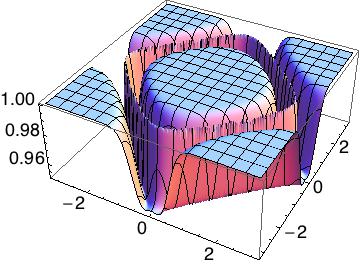}
\par\end{centering}

\caption{\label{fig12} (color online) Total fidelity as a function of momentum. 
Left panel: different temperatures and same finite $\Delta$;
right panel: different temperatures and $\Delta=0$.
}
 
\end{figure}

Finally, in Fig. \ref{fig12} we compare the total fidelity as a function of
momentum for different temperatures. The right panel
corresponds to two quantum states in the normal phase but at different
temperatures. As expected, around the Fermi surface the fidelity decreases
however there is a sharp region where it approaches one. The width of the region
around the Fermi surface is determined by the temperature through the Fermi
function. The sharp maximum corresponds to the point where the two Fermi functions
cross and so it pinpoints the location of the Fermi surface.

\section{Conclusions}

We have introduced and analysed the fidelity spectrum and the fidelity operator
spectrum for partial
states for different systems such as a magnetic impurity in a
conventional superconductor, a $XX$ spin-$1/2$ chain in a transverse
magnetic field and the thermal states of a finite temperature bulk superconductor.

In the first problem we have found that only one charge eigenvalue and one
spin eigenvalue have important changes as the quantum phase transition
induced by the magnetic impurity occurs. The transition is associated with
the capture of one electron by the impurity with a parallel spin.
This feature is clearly seen when we consider two density matrices associated with the same lattice
site (particularly the impurity site) and different but close by spin couplings between
the impurity and the spin density of the electrons. Selecting two density
matrices with the same spin coupling but different lattice sites leads to
a signature of the phase transition that can be seen both from the eigenvalues
associated with lattice sites far from each other and from the same lattice site.

In the spin chain problem we have studied for the first time the block fidelity
and the block fidelity susceptibility and found that the quantum phase transition
that occurs between a $XX$ phase and an Ising like phase is well signaled by
the block fidelity. Both the entanglement spectrum and the fidelity spectrum
do not show any significant features and we analysed the spectra calculating
the moments of the distribution and the R\'enyi entropies. The $S_1$ R\'enyi
entropy associated with the block fidelity shows a distinctive characteristic
away from the critical point.

Finally, in the finite temperature bulk superconductor we showed that in the
superconductor there is a clear gap between the various $k$-eigenvalues, as for
the energy spectrum. In the case of two different density matrices we found
that the effect of temperature is stronger than the difference in the
order parameter distinguishing the normal from the superconducting phase.
Analysing the $k$-fidelity operator spectrum it was clearly seen that the
properties are determined by the structure around the Fermi energy, as
expected. In the case of two density matrices for different temperatures
in the normal phase the fidelity has a sharp maximum at the location of the
Fermi surface, determined by the crossing of the Fermi functions.

We have shown that the fidelity spectrum, which we have introduced, 
can give a more detailed description and characterization of the
phase transitions of many-body quantum systems 
providing complementary information to other techniques.
Therefore, we hope that this can be applied to non-trivial problems where the
traditional Ginzburg-Landau theory with a local order parameter is not known.

\begin{acknowledgments}

NP thanks the project of SQIG at IT, funded by FCT and EU FEDER
projects QSec PTDC/EIA/67661/2006 and QuantPrivTel
PTDC/EEA-TEL/103402/2008, IT Project QuantTel, and Network of
Excellence, Euro-NF. VRV and PDS thank the project PTDC/FIS/64926/2006, of Funda\c
c\~ao para a Ci\^encia e a Tecnologia, Portugal.

\end{acknowledgments}


\begin{thebibliography}{}

\bibitem{Wooters} W. K. Wootters, Phys. Rev. D {\bf 23}, 357 (1981).

\bibitem{Haldane} H. Li and F.D.M. Haldane, Phys. Rev. Lett. {\bf 101}, 010504 (2008). 

\bibitem{Amico} L. Amico, R. Fazio, A. Osterloh, and V. Vedral, Rev. Mod. Phys. {\bf 80},
517 (2008). 

\bibitem{Zanardi} P. Zanardi and N. Paunkovi\'c, Phys. Rev. E {\bf 74}, 031123 (2006).

\bibitem{Gu} S.J. Gu, Int. J. Mod. Phys. B {\bf 24}, 4371 (2010). 

\bibitem{Zhou} H.Q. Zhou, arxiv:0704.2945.

\bibitem{us} N. Paunkovi\'c, P.D. Sacramento, P. Nogueira, V.R. Vieira and V.K. Dugaev,
Phys. Rev. A {\bf 77}, 052302 (2008).

\bibitem{BCS} N. Paunkovi\'c and V.R. Vieira, Phys. Rev. E {\bf 77}, 011129 (2008).

\bibitem{Bernevig} N. Regnault, B. A. Bernevig, and F. D. M. Haldane, Phys. Rev. Lett.
{\bf 103}, 016801 (2009). 

\bibitem{Poilblanc} D. Poilblanc, Phys. Rev. Lett. {\bf 105}, 077202 (2010).

\bibitem{Bernevig2} A. Sterdyniak, N. Regnault, and B. A. Bernevig, 
Phys. Rev. Lett. {\bf 106}, 100405 (2011).

\bibitem{Arovas} R. Thomale, D.P. Arovas and B.A. Bernevig, Phys. Rev. Lett. {\bf 105},
116805 (2010).

\bibitem{Sakurai} A. Sakurai, Prog. Theor. Phys. {\bf 44}, 1472 (1970). 

\bibitem{Balatsky}
A.V. Balatsky, I. Vekhter and J.-X. Zhu, Rev. Mod. Phys. {\bf 78}, 373 (2006).

\bibitem{Entanglement} P.D. Sacramento, P. Nogueira, V.R. Vieira and V.K. Dugaev,
Phys. Rev. B {\bf 76}, 184517 (2007).

\bibitem{Mattis} E. Lieb, T. Schultz and D. Mattis, Annals of Phys. {\bf 16}, 407 (1961).

\bibitem{Barouch} E. Barouch and B. McCoy, Phys. Rev. A {\bf 3}, 786 (1971).

\bibitem{Latorre1} G. Vidal, J.I. Latorre, E. Rico and A. Kitaev, Phys. Rev. Lett.
{\bf 90}, 227902 (2003).

\bibitem{Latorre2} J.I. Latorre, E. Rico and G. Vidal, Quant. Inf. Comput.
{\bf 4}, 48 (2004).

\bibitem{Jin} B.-Q. Jin and V.E. Korepin, J. Stat. Phys. {\bf 116}, 79 (2004).

\bibitem{Giorda} P. Zanardi, M. Cozzini, and P. Giorda, J. Stat. Mech. {\bf 2}, L02002 (2007).

\bibitem{Cozzini} M. Cozzini, P. Giorda, and P. Zanardi, Phys. Rev. B {\bf 75}, 014439 (2007).

\bibitem{Zanardi_g1} P. Zanardi, P. Giorda, and M. Cozzini, Phys. Rev. Lett.
{\bf 99}, 100603 (2007).

\bibitem{Abasto} D. F. Abasto, N. T. Jacobson, and P. Zanardi, Phys. Rev. A {\bf 77}, 022327
(2008).

\bibitem{Garnerone} S. Garnerone, N. T. Jacobson, S. Haas and P. Zanardi,
Phys. Rev. Lett. {\bf 102}, 057205 (2009).

\bibitem{Dubail} J. Dubail and J.-M. St\'ephan, J. Stat. Mech. {\bf 3} L03002 (2011).

\bibitem{Pollmann} F. Pollmann and J.E. Moore, New J. Phys. {\bf 12}, 025006 (2010).

\bibitem{who} A. R. Its and V. E. Korepin, J. Stat. Phys. {\bf 137}, 1014 (2009). 

\bibitem{You} W. L. You, Y. W. Li and S. J. Gu, Phys. Rev. E {\bf 76}, 022101 (2007).

\bibitem{Zanardi_g2} L. Campos Venuti and P. Zanardi, Phys. Rev. Lett. {\bf 99},
095701 (2007).

\end{thebibliography}
\end{document}